\documentclass[12pt]{article}
\usepackage{graphics}
\usepackage{amsthm}
\usepackage{amssymb}
\usepackage{amsmath}
\usepackage{amsfonts}
\usepackage{epic}
\usepackage{hyperref}
\usepackage{epsfig}
\usepackage{bm}
\usepackage{amscd}

\newcommand{\baseRing}[1]{\ensuremath{\mathbb{#1}}}
\newcommand{\Z}{\baseRing{Z}}
\newcommand{\C}{\baseRing{C}}

\newcommand{\R}{\baseRing{R}}

\theoremstyle{plain}

\theoremstyle{definition}

\def\be{\begin{equation}}
\def\ee{\end{equation}} 
\title{The Stoyanovsky-Ribault-Teschner Map and String Scattering Amplitudes}

\author{Gaston Giribet${}^{1,2}$ and Yu Nakayama${}^3$}

\begin{document}
\begin{titlepage}
\begin{flushright}
UT-05-05\\
hep-th/0505203\\
\end{flushright}
\begin{center}
\noindent{{\LARGE{The Stoyanovsky-Ribault-Teschner Map 
}}} 

\noindent{{\LARGE{and String Scattering Amplitudes \\\vspace{0.5cm}
}}} 

\smallskip
\noindent{\large{Gaston Giribet${}^{1,2}$ and Yu Nakayama${}^3$}}
\end{center}
\smallskip
\smallskip
\centerline{${}^1$Department of Physics, Universidad de Buenos Aires, FCEN,}
\centerline{{\it Ciudad Universitaria, 1428. Buenos Aires, Argentina}}
\smallskip
\centerline{${}^2$Department of Physics, Universidad de La Plata, IFLP,}
\centerline{{\it C.C. 67, 1900, La Plata, Argentina}}
\smallskip
\centerline{${}^3$Department of Physics, Faculty of Science,
University of Tokyo} \centerline{\it Hongo 7-3-1, Bunkyo-ku, Tokyo
113-0033, Japan}
\smallskip
\smallskip

\begin{abstract}
Recently, Ribault and Teschner pointed out the existence of a one-to-one correspondence between $N$-point correlation functions for the $SL(2,\C )_k/SU(2)$ WZNW model on the sphere and certain set of $2N-2$-point correlation functions in Liouville field theory. This result is based on a seminal work by Stoyanovsky. Here, we discuss the implications of this correspondence focusing on its application to string theory on curved backgrounds. For instance, we analyze how the divergences corresponding to worldsheet instantons in $AdS_3$ can be understood as arising from the insertion of the {\it dual} screening operator in the Liouville theory side. We also study the pole structure of $N$-point functions in the 2D Euclidean black hole and its holographic meaning in terms of the Little String Theory. This enables us to interpret the correspondence between CFTs as encoding a LSZ-type reduction procedure. Furthermore, we discuss the scattering amplitudes violating the winding number conservation in those backgrounds and provide a formula connecting such amplitudes with observables in Liouville field theory. Finally, we study the WZNW correlation functions in the limit $k \to 0$ and show that, at the point $k=0$, the Stoyanovsky-Ribault-Teschner dictionary turns out to be in agreement with the FZZ conjecture at a particular point of the space of parameters where the Liouville central charge becomes $c_L=-2$. This result makes contact with recent studies on the dynamical tachyon condensation in closed string theory.
\end{abstract}

\end{titlepage}



\newpage

\section{Introduction}\label{sec1}

One of the most useful tools among those employed in the study of non-compact WZNW models is its connection with the Liouville field theory (LFT). This is due to the fact that LFT is the best understood non-rational CFT and hence, any analogy with such theory turns out to be a promising way to extend our understanding of various physics involved in this set of theories.

Examples of such a connection between LFT and the non-compact WZNW models are some identities which relate, as dictionaries, correlation functions of the $SL(2,\C )_k/SU(2)$ WZNW model with those of the Liouville theory. Since this particular WZNW model describes the worldsheet dynamics of strings in Euclidean $AdS_3$, these {\it dictionaries} turn out to be very important in the study of dualities in string theory. In this context, the ultimate goal of this note is to achieve a deeper understanding of the structure of the WZNW model aiming at the $AdS_3/CFT_2$ correspondence.

To be more concise, one such dictionary is the Fateev-Zamolodchikov correspondence \cite{Zamolodchikov:1986bd}, {\it i.e.} a relation between four-point functions in WZNW model and five-point functions in LFT. In the last years this correspondence was extensively used to learn about the $SL(2,\C )_k/SU(2)$ WZNW model. In particular, this led to the proof of its crossing symmetry \cite{Teschner:2001gi}, to the explicit representations of its monodromy \cite{ponsot}, to the understanding of the physical meaning of the KZ logarithmic solutions \cite{Giribet:2004qe}, to the interpretation of certain symmetries in WZNW model, {\it etc}.

More recently, a new dictionary connecting correlation functions in $SL(2,\C )_k/SU(2)$ WZNW model and those in LFT was presented by Ribault and Teschner \cite{rt}, which shows the existence of a one-to-one correspondence between $N$-point correlation functions for the $SL(2,\C)_k/SU(2)$ WZNW model and a certain set of $2N-2$-point correlation functions in Liouville field theory. This correspondence is based on a map between solutions of the BPZ and the KZ systems of partial differential equations that was discovered by Stoyanovsky in \cite{stoyanovsky}.  
The main purpose of this note is to discuss  physical consequences of this map, focusing our attention on applications to the string scattering amplitudes in noncompact curved backgrounds such as Euclidean $AdS_3$ and two dimensional Euclidean black hole.

\subsection{Motivation and main results}
\subsubsection*{The string theory in $AdS_3$ and in the 2D black hole}

String theory in $AdS_3$ has attracted much attention recently because, unlike the higher dimensional $AdS/CFT$ correspondence, the worldsheet CFT has a NS-NS background and can be studied in the ordinary RNS formulation. Our first application of the Stoyanovsky-Ribault-Teschner (SRT) dictionary, therefore, would be to understand  scattering amplitudes in $AdS_3$ from the Liouville theory. The SRT dictionary, connecting observables of both CFTs, provides us a useful tool to reveal several features of the non-compact WZNW model in terms of Liouville theory and possibly {\it vice versa} (the latter is certainly better understood). In particular, this enables us to infer properties of the analytic structure of the scattering amplitudes in string theory in Euclidean $AdS_3$. In this note, after a brief introduction to the CFTs involved and the review of the SRT dictionary, we study the pole structure of the scattering amplitudes and winding number violating process in string theory on $AdS_3$ in terms of the corresponding Liouville correlation functions. From this viewpoint, we will show that the worldsheet instanton effects \cite{mo3} are encoded in the bulk Liouville interaction.

Moreover by gauging a $U(1)$ subgroup of the WZNW model, we can study the correlation functions of $SL(2,\R)_k /U(1)$ coset model by utilizing the SRT map. This coset model was interpreted as a two dimensional black hole \cite{2DBH} and the study of the string theory on this space is, again, an excellent laboratory to enhance our understanding of the stringy physics involving black holes. In addition, $SL(2,\R )_k /U(1)$ coset model is related to the physics near the NS5-brane (so-called Little String Theory \cite{GK}) by the holographic principle. In this note, we will show that the SRT map yields a comprehensive understanding of the pole structure of the correlation functions in LST. Especially, the SRT map separates the LSZ poles and bulk poles naturally in accord with the prediction \cite{Aharony:2004xn} from the consistency with the holographic dual theory.

\subsubsection*{The homogeneity limit of tachyon condensation}

The second motivation to study the SRT map is applications to time-dependent string theories.
Recently, much attention was payed to the time-like versions of noncompact conformal field theories.\footnote{Recently, an interesting idea relating dynamical tachyon condensation phenomena and supercritical string theory was studied in \cite{simeon}.} This is mainly motivated by the fact that these models raise the hope to achieve an exact description of the closed string tachyon condensation. Important steps were made in recent works on these topics by considering the analytic continuation of known non-rational CFTs. In their seminal paper \cite{strominger}, Strominger and Takayanagi initiated a line of investigation mainly based on the computation of correlation functions for these CFTs. In particular, a considerable progress was done in \cite{schomerus}, where Schomerus worked out the exact solution of three-point functions in $c<1$ Liouville theory on the sphere, describing the dynamics of the homogeneous tachyon condensation process in bosonic string theory (see also \cite{tco}). 

In the context of the non-homogeneous tachyon condensation phenomena, Hikida and Takayanagi studied in \cite{takayanagi} the $SL(2,\R )_k$ WZNW model in the range $0 \leq k<2$. By assuming an extrapolation of the FZZ conjecture \cite{FZZ2} (relating the WZNW model and the sine-Liouville field theory) they discussed the observables of the time-like version of this CFT. 
In particular, the authors consider the limit $k\to 0$ of their expressions and compared such limit with the independent computation from the (Wick rotated) Liouville field theory in order to present a consistency check of their results.
Their study of the non-homogeneous tachyon condensation \cite{takayanagi} is currently based on the following two assumptions:

\begin{description}
	\item[i.] The FZZ conjecture is assumed for the range of parameters $0\leq k <2$ in order to connect the time-like sine-Liouville theory and the WZNW model with $k<2$.\footnote{It is well known that the previous consistency checks of this conjecture, based on the functional form of correlation functions in the two CFTs involved, are not valid in this range of parameters (cf. \cite{kkk,fh}). This is due to the substantial change in the functional form of the three-point functions.}
	\item[ii.] It is assumed that the limit $k \to 0$ of the WZNW model ({\it i.e.} of the sine-Liouville theory once assumed {\bf i.}) leads to Liouville theory (times a $U(1)$ free boson). This is, of course, consistent with a heuristic analysis based on the form of the actions of both CFTs. However, in this range of the space of parameters, where the conformal theories become strongly coupled, the analysis simply based on the functional form of the actions is certainly not enough and it is necessary to be proven at the level of the exact correlation functions.\footnote{G.G. thanks V. Schomerus for comments regarding this point.}
\end{description}

Then, the question arises as to whether there is enough evidence to assume the two points listed above. In fact, the geometrical interpretation of both $SL(2,\R )_k/U(1)$ coset model with $k<2$ and Liouville-like models with $\mathrm{Re} (b)=0$ are far from being completely understood. 

One of the purposes of this note is to present some evidence that  FZZ conjecture seems to be consistent in the limit $k \to 0$. To be more specific, we will argue here that any of the two following assertions implies the other one:

\begin{description}
	\item[A.] The $SL(2,\R )_k/U(1)$ coset model is equivalent to the sine-Liouville model (as stated by the FZZ conjecture) at the point $k=0$.
	\item[B.] The Liouville theory with $b^{-2}=-2$ coincides with the sine-Liouville model with $k=0$ (this means that it is proven at the level of $N$-point functions on the sphere).

\end{description}

The first of these items can be thought as the simple extrapolation of the FZZ conjecture. The second assertion, the one regarding the coincidence of Liouville and sine-Liouville models at $b^{-2}=-2$, is supported by the heuristic argument based on the equivalence of both actions at that particular point. At the level of the 2-point function, this equivalence was already proven to be consistent in \cite{takayanagi}.

The strategy of the proof ({\it i.e.} the proof of that {\bf A.} iif {\bf B.}) is rather a simple application of the Stoyanovsky-Ribault-Teschner map. Although the SRT map, in general, gives a correspondence between $N$-point functions in sine-Liouville theory and $2N-2$-point functions in Liouville theory, in this particular limit ($k=0$) we will establish a correspondence between $N$-point functions in both theories by observing a set of remarkable cancellations occurring at the point $k=0$.

In this way, we will discuss in this note the limit $k\to 0$ of the $SL(2,\R )_k/U(1)$ coset model and prove that, if one assumes the FZZ conjecture, the $N$-point functions in both Liouville theory (times a $U(1)$ free boson) and those in sine-Liouville field theory actually coincide (at quantum level). This represents a proof of the homogeneity limit considered in \cite{takayanagi}. Conversely, if one feels that the equivalence between Liouville and sine-Liouville conformal models at $b^{-2}=-2$ ({\it i.e.} $k=0$) is guaranteed {\it ab initio} by the coincidence of both actions at that point (thus, the check of the agreement of correlation functions would be a trivial consequence of that), then our result could be regarded as a consistency check of the FZZ conjecture at the particular point $k=0$.


\subsection{Overview} 

We will dedicate the following section \ref{sec2} to preliminary points. After a brief introduction to the conformal field theories involved, we review the dictionary between WZNW correlation functions and those of LFT. In section \ref{sec3} we will discuss the applications of such dictionary to the study of string scattering amplitudes in noncompact curved backgrounds. In subsection \ref{sec3-1} we will analyze the pole structure of $SL(2,\C )_k/SU(2)$ WZNW correlation functions with the intention to identify those poles which were interpreted in \cite{mo3} as worldsheet instantons in $AdS_3$. In subsection \ref{sec3-2} we continue our analysis of the pole structure and study the separation of the LSZ poles from the bulk poles in the 2D black hole. This is of importance in the context of holography and Little String Theory. In subsection \ref{sec3-3}, the correlation functions that represent violating winding processes are studied. We present a closed formula which extends the Stoyanovsky-Ribault-Teschner dictionary in order to include winding violating amplitudes. Our generalization is based on the FZZ prescription for the computation of such amplitudes in the Euclidean black hole. Then, we study the limit $k\to 0$ of sine-Liouville correlation functions and discuss its relation with the FZZ conjecture. The analysis of this limit, whose result is presented in subsection \ref{sec3-4}, is related to recent studies on dynamical tachyon condensation in closed bosonic string theory. In section \ref{sec4}, we present discussions and conclusions.

\section{Preliminary: Conformal Field Theory}\label{sec2}

In this section we will describe some useful aspects of both the $SL(2,\C )_k/SU(2)$ WZNW model and Liouville field theory and establish the correspondence between the correlation functions in these theories. Let us begin with the WZNW theory.

\subsection{The $SL(2,\C )_k/SU(2)$ WZNW model}\label{sec2-1}
Besides the Liouville model, there are no many non-rational conformal theories for which our understanding can be considered satisfactory. However, the example  which is clearly closest to that situation is the WZNW model formulated on $SL(2,\C )_k/SU(2)$, where $k$ denotes the level of the current algebra. Among other applications, this field theory represents the non-linear sigma-model describing the string theory in Euclidean $AdS_3$ space $(\sim H^+_3)$ in presence of NS-NS magnetic field.

The action of this theory is 
\begin{equation}
S_{WZNW} =\frac {1}{2\pi}\int d^{2}z\left( \partial \phi \bar{\partial}\phi - \frac {\sqrt{2}}{4}\frac{1}{\sqrt {k-2}} R \phi +
\partial \bar {\gamma} \bar {\partial } \gamma e^{-\sqrt {\frac {2}{k-2}}\phi }\right)  \ ,\label{swznw}
\end{equation}
where $R$ is the worldsheet scalar curvature, 
and this action defines a non-compact conformal field theory with a central charge $c=3+\frac{6}{k-2}$.\footnote{We will use $\alpha'=2$ notation throughout this paper.} Alternative way to describe the system is to utilize a group theoretical representation of the vertex operators of this theory: These are often denoted as $\Phi _j$ and are the operators that create states of the $\hat {sl(2)}_k$ representations from the invariant vacuum $\left| 0 \right\rangle $. For convenience, we will employ two different basis to represent these operators: First, let us introduce the often called $(m,\bar m)$-basis, creating the vectors $\left| \Phi \right\rangle = \left| j,m \right\rangle \otimes \ \left| j,\bar m \right\rangle $ of the representations of the affine Kac-Moody algebra $\hat {sl(2)}_k \otimes \hat {sl(2)}_k $ by means of the application $\lim_{z,\bar{z} \to 0}\Phi _{j,m,\bar m } (z,\bar z) \left| 0 \right\rangle = \left| j,m \right\rangle$. These states have conformal dimension $h_j$ given by
\[
h_j = -\frac {j(j+1)}{k-2} \ .
\]

In the context of string theory in $AdS_3$, where the WZNW model plays a central role, the spectrum is constructed in terms of a subset of representations; see \cite{mo1}. Local operators $\Phi _{j,m,\bar m }(z)$ are also associated to differential functions $\Phi _j (x|z)$ on the hyperbolic upper half hyperplane $H^+_3$. These functions are given by the following decomposition
\begin{equation}
\Phi_{j,m,\bar m } (z) = \int d^2x \ \Phi_{j} (x|z) \ x^{-1-j+m} \bar{x}^{-1-j+\bar m} \label{vertex} \ ,
\end{equation}
which defines what we will call the $(x,\bar x)$-basis. Because of the geometrical reason (or the locality of the vertex operator), $(m,\bar{m})$ has a quantization condition:
\begin{equation}
m = \frac{n+k\omega}{2}, \ \ \bar{m} = \frac{-n+k\omega}{2}, \ \ n\in \Z , \ \ \omega \in \Z \ . \label{quant}
\end{equation}
For these $(x,\bar{x})$-basis vertex operators, we will make use of the harmonic function realization that connects with the WZNW action \eqref{swznw}; namely
\begin{equation}
\Phi_j (x |z) = \frac {2j+1}{\pi} e^{2j\psi(x|z)} \ \ \ \ , \ \ \psi (x|z) = \log (|\gamma -x|^2 e^{\phi}+ e^{-\phi}) \ ,
\end{equation}
which holds for the classical theory. This representation in terms of variables $x$ and $\bar x$ is quite useful in the study of the $AdS_3/CFT_2$ correspondence due to the fact that these complex variables $(x,\bar x)$ turn out to correspond to the coordinates on the boundary of the $AdS_3$, where the dual 2D CFT is defined.

After introducing quantum (finite $k$ or finite $\alpha'$) corrections, it is feasible to show that the large $\phi $ behavior of the vertex operators is given by
\begin{equation}
\Phi_{j,m,\bar m } (z) \sim \gamma ^{-1-j+m} \bar{\gamma } ^{-1-j+\bar m}  e^{-\sqrt {\frac {2}{k-2}}(j+1)\phi} + {\mathcal R}_k (j, m) \ \gamma ^{j+m} \bar {\gamma }^{j+\bar m} e^{\sqrt {\frac {2}{k-2}}j\phi} \ ,
\end{equation}
where
\begin{equation}
{\mathcal R}_k (j, m) = - \left( \pi M^{-1} \frac {\Gamma (\frac{k-3}{k-2})}{\Gamma (\frac{k-1}{k-2})}\right) ^{2j+1} \frac {\Gamma (2j+1) \Gamma (1+\frac{2j+1}{k-2})\Gamma(-j+m) \Gamma (-j-\bar{m})}{\Gamma (-2j-1) \Gamma (1-\frac{2j+1}{k-2}) \Gamma (j+1+m)\Gamma(j+1-\bar{m})} \ .\label{refwz}
\end{equation}
In the interpretation of the $SL(2,\R)_k/U(1)$ model as a two-dimensional black hole, the parameter $M$ precisely represents the black hole mass.

Correlation functions in the WZNW model are known to satisfy the Knizhnik-Zamolodchikov (KZ) system of differential equations. In the case of $SL(2,\R)_k$ WZNW model, the correlation functions involving the primaries (\ref{vertex}) are defined as
\[
{\mathcal A}^{WZNW}_{j_{1},...j_{N}; m_1,m_2, ...j_N} = \langle \Phi _{j_1, m_1, \bar {m}_1} (z_1) ... \Phi _{j_N, m_N, \bar{m}_N} (z_N) \rangle 
\]
and were first computed by Becker and Becker in \cite{BB} by using free field techniques. This realization involves the insertion of $s$ screening charges $\sim e^{-\sqrt{\frac{2}{k-2}}\phi}$, being
\begin{equation}
s=1-N-\sum _{i=1}^N j_i  \label{pocho} 
\end{equation}
as can be obtained by integrating the zero-mode of the field $\phi $ in the correlation functions \cite{BB,gn3}. The explicit expression of two and three-point correlation functions (for $k>2$) are known and reviewed in the following. 

The two-point function takes the form
\begin{eqnarray}
{\mathcal A}^{WZNW}_{j_1,j_2}(x_1,x_2)=\left| z_{1}-z_{2}\right| ^{-4h_{j_1}} \left( \left|
x_{1}-x_{2}\right| ^{-4j_1} B(j_1) \delta (j_1-j_2) + \delta ^{(2)} (x_1-x_2) \delta (j_1+j_2-1) \right) \ , \nonumber \\ 
\label{cacarulo1}
\end{eqnarray}
where
\begin{eqnarray}
B(j)= \frac {k-2}{\pi } \left( \pi M^{-1} \frac{\Gamma \left( \frac{k-3}{k-2} \right) }{%
\Gamma \left(\frac{k-1}{k-2} \right) }\right) ^{2j+1} \frac {\Gamma ( 1+\frac {2j+1}{k-2})}{\Gamma ( -\frac {2j+1}{k-2})} \ .
\end{eqnarray}
This is the $SL(2,\R )_k$ reflection coefficient and satisfies\footnote{In these equations we consider $M=1$ for simplicity.}
\begin{eqnarray}
B(j)B(-1-j) &=& - \frac {1}{\pi ^2} (2j +1)^2 \\
B(j)B(-\frac k2-j) &=& \frac {1}{\pi ^2} B( 1-\frac k2) \ .
\end{eqnarray}

On the other hand, we have the following expression for the three-point function
\begin{eqnarray}
{\mathcal A}^{WZNW}_{j_{1},j_{2},j_{3};m_1,m_2,m_3} = \prod _{a=1}^3 \int d^2x_a x_a^{-1-j_a+m_a}\bar {x}_a^{-1-j_a+\bar {m}_a} \ {\mathcal A}^{WZNW}_{j_{1},j_{2},j_{3}}(x_1,x_2,x_3)
\end{eqnarray}
with
\begin{eqnarray}
{\mathcal A}^{WZNW}_{j_{1},j_{2},j_{3}}(x_1,x_2,x_3)=\prod_{a<b}\left| z_{a}-z_{b}\right|
^{2(h_{j_c} -h_{j_a} -h_{j_b} )}\left| x_{a}-x_{b}\right| ^{2(j_c -j_a -j_b )} C^H (j_{1},j_{2},j_{3})  \label{cacarulo2} \ ,
\end{eqnarray}
where $a,c,b \in \{1,2,3 \}$ and
\[
C^{H}(j_{1},j_{2},j_{3})= \frac {k-2}{2\pi ^3} \left( \pi M^{-1} \frac{\Gamma \left(\frac {k-3}{k-2} \right) }{\Gamma \left( \frac {k-1}{k-2} \right) }\right) ^{2+j_1+j_2+j_3} \frac{G_k (1+j_{1}+j_{2}+j_{3})}{G_k (-1)} \prod _{a=2}^{4}   \frac {G_k (-2j_a+j_{1}+j_{2}+j_{3})}{G_k (2j_{a}+1)} \ .
\]
In this expression, $G_k (x)$ are special functions that can be written in terms of $\Gamma _2 (x|1,y)$ Barnes functions; namely
\[
G_{k}(x)\equiv (k-2)^{\frac{x(k-1-x)}{2k-4}}\Gamma _{2}(-x\mid 1,k-2)\Gamma _{2}(k-1+x\mid 1,k-2) 
\]
with
\[
\log (\Gamma _{2}(x\mid 1,y))\equiv \lim_{\varepsilon \rightarrow 0}\frac {\partial}{
\partial \varepsilon } \left( \sum_{n=0}^{\infty }\sum_{m=0}^{\infty
}(x+n+my)^{-\varepsilon } + \sum_{n=0}^{\infty }\sum_{m=0}^{\infty
}(\delta _{n,0}\delta _{m,0}-1) (n+my)^{-\varepsilon }\right) \ .
\]
Alternatively, these functions can be written in terms of the $\Upsilon _b (x)$ functions introduced in Liouville literature by Zamolodchikov and Zamolodchikov \cite{Zamolodchikov:1995aa}, see below.

The physical information of the formula (\ref{cacarulo2}) is encoded in the analytic properties of $G$ functions; for our purpose it is enough to mention that $G_k (x)$ presents poles in the set $x \in \Z _{<0}+ \Z _{<0} (k-2)$ and $x \in \Z _{\geq 0}+ \Z _{\geq 0} (k-2)$.

The functions $C^{H}(j_1,j_2,j_3)$ are denoted as structure constants and are clearly invariant under any permutation of the set $\{ j_{1},j_{2},j_{3} \}$. Moreover, these satisfy the following functional properties:
\begin{eqnarray}
\lim _{j_3 \to 0} C^{H}(j_1,j_2,j_3) &=& B(j) \delta (j_1-j_2) \label{catorce} \\
\lim _{j_3 \to -\frac k2}C^{H}(j_1,j_2,j_3) &=& \frac {1}{\pi } B( 1-\frac k2) \delta (j_1+j_2+\frac k2)  \label{quince} \ ,
\end{eqnarray}
which realize the relation between the structure constants and reflection coefficients.

These correlation functions were studied in detail in the context of the $SL(2,\C )_k/SU(2)$ WZNW model in \cite{T1,T2,T3}, where alternative derivations were presented. Regarding the applications to string theory on $AdS_3$, correlation functions describing scattering amplitudes in this space were carefully analyzed by Maldacena and Ooguri in \cite{mo3}, proposing an analytic extension to the case of $SL(2,\R )$ which describes the Lorentzian $AdS_{2+1}$; see also \cite{gn3} and references therein.

Now, let us move on to Liouville field theory.

\subsection{The Liouville field theory}\label{sec2-2}

Liouville field theory is one of the best understood non-rational conformal field theories and its importance in the context of string theory and 2D quantum gravity has made this model one of the central subjects of investigation in mathematical physics. Due to this, in the last ten years we have made a substantial improvement in  understanding the structure of the correlation functions on the sphere (see \cite{Teschner:2001rv,Nakayama:2004vk} for reviews).

The action of the quantum Liouville theory is written as follows
\begin{equation}
S_L =\frac {1}{2\pi} \int d^{2}z\left( \partial \varphi\bar{\partial}\varphi + \frac{\sqrt{2}}{4}QR\varphi +2\pi\mu _+
e^{\sqrt {2}b\varphi }\right)  \ ,\label{sliouville}
\end{equation}
where $\mu _+$ is the Liouville cosmological constant and $R$ is the two-dimensional scalar curvature. The theory has a central charge $c_L =1+6Q^{2}$, where the background charge is given by $Q=(b+b^{-1})$.

The cosmological constant term ({\it i.e.} the last term in (\ref{sliouville})) corresponds to one of the two exponential screening operators of this theory. Indeed, in Liouville field theory, there are two different screening operators:
\[
S_{\pm }=\mu _{\pm }\int d^2zV_{b^{\pm 1}}(z)=\mu _{\pm} \int d^2 z e^{\sqrt{2}b^{\pm 1}\varphi (z)} \ ,
\]
where the dual cosmological constant $\mu_-$ is given by 
\begin{equation}
\mu _- = \frac {\Gamma (1-b^{-2})}{\pi \Gamma (b^{-2})} \left( \pi \frac {\Gamma (b^2)}{\Gamma (1-b^2)}\mu _+ \right) ^{b^{-2}}, \label{cacho7}
\end{equation} 
see \cite{Teschner:2001rv,Zamolodchikov:1995aa,gn3}.

The exponential primary fields of the following form turn out to be very important objects in Liouville field theory
\[
V_{\alpha }(z)=e^{\sqrt{2}\alpha \varphi (z)} \ .
\]
These have conformal dimension $\Delta _{\alpha }=\alpha (Q-\alpha )$. Notice that $\Delta _{\alpha}$ is invariant under the reflection $\alpha \rightarrow Q-\alpha$, and corresponding states would be related by such a conjugation since one can consider the composed vertex operator including both contributions $V_{\alpha }$ and $V_{Q- \alpha }$ as
\begin{equation}
{\mathcal V}_{\alpha }(z)= V_{\alpha }(z) + R_b ({\alpha}) V_{Q- \alpha }(z)  \label{tyt} \ ,
\end{equation}
where $R_b({\alpha })$ is the Liouville reflection coefficient ({\it i.e.} given by Liouville two-point function). The explicit expression of the reflection coefficient takes the form 
\[
R_b ({\alpha })=- \left( \pi \mu _+  \frac{\Gamma (b^2)}{\Gamma (1-b^2)} \right) ^{1-2\alpha b^{-1}+b^{-2 }} \frac {\Gamma (2b\alpha -b^2 ) \Gamma (2b^{-1}\alpha -b^{-2})}{\Gamma (2-2b\alpha +b^2 ) \Gamma (2-2b^{-1}\alpha +b^{-2 })} \ .
\]
Thus we have an operator identification
\begin{equation}
{\mathcal V}_{Q-\alpha }(z) \equiv R_{b} ({Q-\alpha }) {\mathcal V}_{\alpha }(z) \ .
\end{equation}

In the large $\varphi $ region, the correlation functions are controlled by the expectation values of local operators $V_{\alpha } (z)$ and are written as
\[
{\mathcal A}^{L}_{\alpha _{1},...\alpha _{N}} = \ \langle V_{\alpha 
_{1}}(z_1)V_{\alpha _{2}}(z_2)...V_{\alpha _{N}}(z_N)\rangle  \ .
\]
The basic observation to derive Liouville correlation functions is to notice that the residue of the poles in the correlation functions should be given by the following screening integral
\begin{equation}
{\mathcal A}_{\alpha _{1},...\alpha _{N}}= \frac {\mu _+ ^{ n_+ } \mu _- ^{ n_- }}{n_+!n_-!} \langle \prod _{i =1} ^{N} e^{\sqrt{2}\alpha _{i }\varphi
(z_{i })} \prod^{n_{-}}_{r=1} \int 
d^2v_{r} e^{\sqrt{2}b^{-1} \varphi (v_r)}\prod^{n_{+}}_{r=1} \int d^2w_{r} e^{\sqrt{2}b 
\varphi (w_r)}\rangle  \ ,  \label{repint}
\end{equation}
where $n_{\pm }$ refers to the amount of screening operators of the type $S_{\pm }$ required to satisfy the charge symmetry condition coming from the integration over the zero-mode of $\varphi $ field (and treat $\varphi$ as if it were a free field);
\begin{equation}
\sum _{i =1} ^{N} \alpha _{i }+n_{+}b+n_{-}b^{-1}=Q \ .   \label{clot}
\end{equation}
In the full correlation functions, we expect poles such as 
$\Gamma(Q-\sum _{i =1} ^{N} \alpha _{i }+n_{+}b+n_{-}b^{-1})$.

The explicit form of the two and three-point correlation functions for real values of $b$ are known and we write them here to archive a self-contained presentation. Let us start with the two point-function, which is just given in terms of the reflection coefficient as
\begin{equation}
{\mathcal A}^L_{\alpha _1, \alpha _2} = |z_2-z_1|^{-4\Delta _{\alpha _1}} \left(R_b (\alpha _1)\delta(\alpha_1-\alpha_2) + \delta(\alpha_1+\alpha_2-Q) \right)\ .
\end{equation}

On the other hand, the three-point function is given by
\begin{equation}
{\mathcal A}^L_{\alpha _1, \alpha _2, \alpha _3} = |z_1 - z_2|^{2(\Delta _{\alpha _3} -\Delta _{\alpha _2}-\Delta _{\alpha _1})}           |z_1 - z_3|^{2(\Delta _{\alpha _2} -\Delta _{\alpha _3}-\Delta _{\alpha _1})}          |z_3 - z_2|^{2(\Delta _{\alpha _1} -\Delta _{\alpha _2}-\Delta _{\alpha _3})}    C^L (\alpha _1, \alpha _2, \alpha _3)
\end{equation}
with the structure function $C^L(\alpha _1, \alpha _2, \alpha _3)$ defined as follows
\begin{equation}
C^L(\alpha _1, \alpha _2, \alpha _3) = \left( \pi \mu _+ \frac {\Gamma (1+b^2)}{\Gamma (1-b^2)} b^{-2b^2}\right) ^{1+b^{-2}-b^{-1}\sum _{n=1}^3\alpha _n} \frac {\Upsilon _b' (0)}{\Upsilon _b (\sum _{t=1}^3 \alpha _t -Q)}  \prod _{i=1}^3 \frac {\Upsilon _b (2\alpha _i)}{\Upsilon _b (\sum _{l=1}^3 \alpha _l -2 \alpha _i)}  \label{C} \ .
\end{equation}
Here the special functions $\Upsilon _b(x)$ can be written in terms of the special functions $G_{k}(x)$ introduced before by means of the equivalence 
\begin{equation}
\Upsilon _b ^{-1}(-bx)=b^{b^{2}x^{2}+(1+b^{2})x}G_{b^{-2}+2}(x) \ ,  \label{ide}
\end{equation}
where $b^{-2} = k-2$.

The two and three-point functions written above were originally computed by Dorn-Otto and Zamolodchikov brothers in \cite{Dorn:1994xn,Zamolodchikov:1995aa}, where they derived them as interpolations of the screening integral \eqref{repint}. Later, Teschner \cite{Teschner:1995yf} reobtained this expression from the consistency condition of the crossing symmetry of a certain four-point function with one degenerate operator insertion (conformal bootstrap approach).

In Liouville field theory, there exist particular representations of Virasoro algebra which correspond to degenerate states. The use of them turned out to be a crucial tool in our understanding of the spectrum and the set of observables of this non-rational theory. This is basically due to the fact that correlation functions involving such states are known to satisfy BPZ partial differential equations. The values of momenta corresponding to those representations are given by
\[
\alpha = \alpha _{m,n} = \frac{1-m}{2}b + \frac{1-n}{2}b^{-1} \ , 
\]
where $(m,n)$ is any pair of positive integers. The state with momentum $\alpha = \alpha _{1,2}$ will play a distinguishing role in what follows.

\subsection{The Stoyanovsky-Ribault-Teschner dictionary}\label{sec2-3}


Based on a relation between the BPZ and the KZ systems of partial differential equations, Stoyanovsky described in \cite{stoyanovsky} a map between correlation functions between $SU(2)$ WZNW models and minimal models. On this basis, Ribault and Teschner recently presented a precise formula connecting correlation functions in $SL(2,\C )_k/SU(2)$ WZNW theory and those in Liouville theory \cite{rt}. The formula, which we will call Stoyanovsky-Ribault-Teschner (SRT) dictionary, maps $N$-point functions in WZNW theory to $2N-2$-point functions in Liouville theory. Explicitly,
\begin{eqnarray}
\langle \prod _{i=1}^N \Phi _{j_i,m _i,\bar {m}_i}(z_i)  \rangle &=&  \prod _{i=1}^{N}N^{j_i}_{m_i, \bar {m} _i}  \prod_{r=N+1}^{2N-2} \int d^2z_r \ F_k (z_1, ... z_{2N-2}) \  \langle \prod _{t=1}^{2N-2} V_{\alpha _t} (z_t) \rangle \ , \label{rt}
\end{eqnarray}
where 
\begin{eqnarray}
F_k (z_1, ...z_N, z_{N+1}, ... z_{2N-2}) &=&  \frac {2\pi ^3b}{\pi ^{2N}} (\mu _+ \pi ^2 /Mb^{2})^{b^{-1}(\alpha _1+...\alpha _N-\frac {N}{2} b^{-1}) -1} \times \label{cacho8} \\
&& \times \frac {\prod_{1\leq r<l}^{N}(z_r-z_l)^{m_r+m_l+k/2}(\bar{z}_r-\bar{z}_l)^{\bar{m}_r+\bar{m}_l+k/2} \prod_{N<r<l}^{2N-2}|z_r-z_l|^{k}}{\prod _{t=1}^{N}\prod_{r=N+1}^{2N-2} (z_r-z_t)^{m_t+k/2}(\bar{z}_r-\bar{z}_t)^{\bar{m}_t+k/2} }  \ ,\nonumber \end{eqnarray}
and
\begin{eqnarray}
N^{j}_{m,\bar m} = -\frac {\sin (\pi(2j-m-\bar m))\Gamma (-j+\bar m)\Gamma (-j+m)}{2\pi  \ \cos (\pi (j-m))} = \frac{\Gamma(-j+m)}{\Gamma(1+j-\bar{m})} \ .
\end{eqnarray}
The map of the vertex operators is given by 
\begin{eqnarray}
\alpha _r = bj_r+b+b^{-1}/2 \ \ \ , \ \ 1\leq &r& \leq N \\ \label{mapl}
\alpha _r = \alpha _{1,2} = -b^{-1}/2 \ \ \ , \ \ N < &r& \leq 2N-2 \ . 
\end{eqnarray}
In the expression above we have also identified the parameters $b^{-2} = k-2$ and taken into account the following constraints\footnote{Later we will discuss the winding number violating correlation functions and relax (\ref{mapl}).}
\begin{equation}
\sum _{i=1}^Nm _i=\sum _{i=1}^N\bar {m} _i=0 \ .
\end{equation}
Expression (\ref{rt}) can be also written down in terms of the variables $(\mu , \bar \mu)$ introduced in \cite{rt} as a Laplace-Fourier transform of $(x,\bar{x})$-basis:
\begin{eqnarray}
\Phi_j(\mu|z) = \frac{1}{\pi}|\mu|^{2j+2}\int d^2x e^{\mu x-\bar{\mu}\bar{x}} \Phi_j(x|z) \ . \label{xbas}
\end{eqnarray}
The correlation functions in this basis are proportional to the factor 
\begin{equation}
\propto |\Theta _N (\mu | z)^{-1} \sum _{i=1}^N \mu_i z_i|^{-2-4b^2} \ ,  \label{caballo}
\end{equation}
where the function $\Theta _N(\mu | z)$ and the variables $\mu$ and $\bar \mu$ are defined by the following relations
\begin{equation}
|\Theta _N(\mu | z)| ^2 = |\sum_{i=1}^N z_i \ \mu_i |^2  \prod _{i<r\leq N}^{N-1,N} |z_i - z_r|^{b^{-2}}   \prod _{N<i<r}^{2N-3,2N-2} |z_i-z_r|^{b^{-2}}    \prod _{1\leq i}^{N} \  \prod _{N<r}^{2N-2} |z_i-z_r|^{-b^{-2}}  \label{theta}
\end{equation}
and
\begin{equation}
\sum _{i=1}^N\mu _i=0 \ ; \ \ \ \sum_{i=1}^{N} \mu_i (t-z_i)^{-1} = \prod _{1\leq r}^{N} (t-z_r)^{-1} \prod_{N<r}^{2N-2} (t-z_r) \  \sum_{i=1}^{N} z_ i \ \mu_i
\end{equation}
and, of course, the corresponding anti-holomorphic part (see \cite{rt} for the details).

\subsection{Some remarks on SRT correspondence}\label{sec2-4}
Before concluding this section and move on to our main study of the SRT correspondence (\ref{rt}) in the applications to string theory formulated on $AdS_3$ and on the 2D black hole background and homogeneous tachyon condensation, we would like to briefly make several remarks on the SRT correspondence purely from the CFT point of view, which is of its own interest and also is of importance in our applications.

\begin{itemize}
	\item The original derivation of the SRT map implicitly assumes that $\alpha = \frac{Q}{2} + iP$ and $j = -\frac{1}{2}+ ip$ with real $P$ and $p$  (continuous series).\footnote{This is because in more general amplitudes, we have to specify the contour of the intermediate channel when one applies the OPE in order to accommodate the contribution from discrete states.} In the following applications, we assume that the SRT map can be applied to real $\alpha$ and $j$ by postulating sufficient analyticity of the correlation functions in both theory (defined as an analytic continuation). Such an analytic continuation is believed to hold for higher point functions and shown explicitly in lower (two and three) point functions.
	\item The relation between the quantum numbers $\alpha $ and $j$ is such that the Liouville reflection symmetry $\alpha \to Q-\alpha $ corresponds to the Weyl symmetry transformation $j \to -1-j$ of $SL(2,\R )$ representations. Moreover, (\ref{mapl}) also shows that the degenerate representations $\alpha _{m,n}$ of the Liouville spectrum (corresponding to non-normalizable states) are mapped to the $j^-_{m,n}$ representations of the Kac-Kazhdan series of the $\hat {sl(2)} _k$ algebra (we will return to this point in the conclusion).
	\item It is also worthwhile mentioning that the Liouville theory possesses a self-duality under the interchange $b \to b^{-1}$. However, this symmetry, under the SRT map, apparently is not present in the $SL(2,\R)_k$ WZNW model, whose central charge $c=3+\frac{6}{k-2}$ is asymmetric under $k-2 \to 1/(k-2)$ unlike the Liouville theory ($c = 1 + 6(b+b^{-1})^2$). This difference is encoded in the fact that the function $F_k(z_1, ...z_{2N-2})$ in (\ref{rt}) explicitly depends on $k$. However, it is also true that under the transformation $k-2 \to 1/(k-2)$ the WZNW correlation functions transform in such a way that a relation of the form (\ref{cacho7}) also holds, considering $M$ instead of $\mu $ (see \cite{gn3} for the details). This is consistent with the fact that $\mu _+$ and $M$ appear together in the KPZ scale factor in (\ref{cacho8}).
	\item Related to this point, it is now widely believed there is a phase transition (known as the String/Black Hole transition \cite{Giveon:2005mi}) at $k=3$ of the $SL(2,\R)_k$ WZNW model. Since the Liouville part of the SRT map is invariant under $b\to b^{-1}$, the whole information of the phase transition should be encoded in $F_k(z_1, ...z_{2N-2})$. It would be interesting to study this aspects further, but beyond the scope of this note (see also \cite{Nakayama:2004ge} for an explicit realization of this transition in the open string sector).

\end{itemize}

\section {Applications to String Theory}\label{sec3}

The aim of the following subsections is to investigate how we can extract physically useful information about the functional properties of string scattering amplitudes from the SRT map.

\subsection{The instantons in Euclidean $AdS_3$}\label{sec3-1}

As a first simple application, we study the pole structure of $N$-point scattering amplitudes on $AdS_3$. The analytic structure of string amplitudes in Euclidean $AdS_3$ was extensively studied by Maldacena and Ooguri in \cite{mo3}, where it was shown that the four-point scattering amplitude in this space presents additional poles with respect to the analytic structure one could naively expect {\it a priori}. These additional  poles  are located in the middle of the moduli space when\footnote{where $x$ and $z$ refer to the cross-ratios of both $x_i$ and $z_i$ insertions in the case of four-point functions written in terms of the $(x,\bar x)$-basis. Namely, $x=\frac{(x_1-x_3)(x_4-x_2)}{(x_4-x_1)(x_2-x_3)}$ and $z= \frac{(z_1-z_3)(z_4-z_2)}{(z_4-z_1)(z_2-z_3)} $.} $x=z$. By means of semiclassical arguments, these poles were interpreted in \cite{mo3} as due to instantonic contributions in the worldsheet theory and happen when a holomorphic map $\gamma (z) \to z$ does exist (typically, one can consider the case $\gamma (z)= z ^{\omega} $). In that case, the third term in action (\ref{swznw}) vanishes and no potential that prevents the worldsheet from expanding to the boundary is present. From the viewpoint of the dual boundary conformal field theory, these non-perturbative effects (finite $k$-effects) are seen as non-local phenomena. 

In \cite{mo3}, it was also inferred (from the semiclassical analysis) that in a generic $N$-point function on the sphere, after integrating over $z_i$, one finds those points in the $x$-plane where the holomorphic map is possible and, thus, one would expect poles at
\begin{equation}
k+N-3+\sum _{i=1}^{N} j_i=0 \ \ ,  \label{asterisco}
\end{equation}
corresponding to instantons wrapping $\omega =1 $ times around themselves and expanding to the $S^2$ that is the boundary of $AdS_3$. Here, we argue that these ``expected'' poles actually occur in $AdS_3$ $N$-point amplitudes. By using the Stoyanovsky-Ribault-Teschner dictionary, it is easy to show that the $N$-point functions in $SL(2,\C)_k/SU(2)$ WZNW model actually present the pole structure that Maldacena and Ooguri predicted. They simply appear from the bulk poles coming from the corresponding $2N-2$-point Liouville correlation functions. The mechanism creating these poles in the Liouville side is explained as follows: While in the $AdS_3$ picture they correspond to worldsheet configurations expanding to the boundary ($\phi \to \infty$) with no potential preventing that, from the viewpoint of Liouville theory these poles arise due to the non-compactness of the target space and the integration over the zero-mode $\varphi _0$ on it when the configuration is such that an integer amount of screening charges $S_-$ is inserted.\footnote{G.G. thanks J.M. Maldacena for earlier conversations on this point. Actually, Maldacena suggested before \cite{rt} that it could be possible to find a connection between the instantonic poles in $AdS_3$ and the {\it dual} screening operator. This seems to be the explicit realization.}

To be precise, let us consider the integral representation of the Liouville correlation functions (a Coulomb gas-type representation); for instance, let us consider the case where only screenings of the kind $S_- \sim \int d^2 v e^{\sqrt{2} b^{-1}\varphi (v)}$ are present in (\ref{repint}). Then, when computing the Liouville correlation functions, one picks up a factor $\Gamma (-n_-)$ coming from the integration over the zero-mode of the Liouville field when $n_-$ screening charges are inserted according to (\ref{clot}). Consequently, the correlation functions present poles for those configurations such that $n_-\in \Z _{\geq 0}$; see for instance \cite{dFK}. In particular, the case $n_-=1$ precisely corresponds to the Maldacena-Ooguri condition (\ref{asterisco}). For a generic non-negative integer $n_-$, the pole we get is the one corresponding to a worldsheet instanton wrapping $\omega = n_-$ times in $AdS_3$. If we consider the case $n_-=1$ in (\ref{clot}), bulk poles for the 4-point functions also arise for $n_+\geq 0$; these correspond to the fact that the monodromy invariant solution of the $SL(2,\R )_k$ Knizhnik-Zamolodchikov equation have a contribution of the form $\sim |z-x|^{-2n_+}$. The poles at integer $n_{\pm}$ appear here because of the integration over the variables $(x,\bar x)$, which was already performed when writing the correlation functions in terms of the $(m,\bar m)$-basis.

According to the spacetime picture, the poles due to the function $\Gamma (-n_+)$ that stands in Liouville correlation functions are interpreted as contributions due to tachyons scattering in the bulk with $n_+$ zero-momentum tachyons of the background, which correspond to the interaction with the microscopic composites of the Liouville potential \cite{dFK}. In the case of $n_-$ screenings of the type $S_-$ (the dual interaction term) this would correspond to instantonic contributions. In more general situations, the both effects contribute as a bulk pole.

\subsection{Separation of LSZ poles from bulk poles}\label{sec3-2}

In \cite{Aharony:2004xn}, it was claimed that the pole structure of the $SL(2,\R )_k/U(1)$ theory yields a holographic meaning in the context of the dual Little String Theory (LST). Roughly speaking, each vertex operator corresponding to operators in the dual theory which creates a one particle state gives  LSZ-like poles. On the other hand, there are other kinds of poles coming from the bulk interaction related to the noncompact cigar geometry (known as bulk poles); this was discussed in the context of SRT map in the last subsection. As we will see in this subsection, the separation of LSZ poles from bulk poles studied in \cite{Aharony:2004xn} can be shown manifestly by using the SRT map.\footnote{It was explicitly discussed in two and three point functions; we will show that the structure holds for arbitrary $N$-point functions.}

The SRT map, when we write in the $m$ basis, assigns a normalization factor $N^j_{m,\bar{m}}$ to each vertex of the WZNW model. This factor is based on the group theoretical integral, which transforms the $\mu$-basis to the $m$-basis of the $SL(2,\R )$ representation. On the other hand, in \cite{Aharony:2004xn} from the dual LST description, it was suggested that  every vertex operator representing the off-shell source for the Green function should possess a definite pole structure predicted from the LSZ reduction.

{}From the usual gauge/string correspondence, the on-shell string amplitudes calculate off-shell Green functions of the gauge theory (LST in this case). Therefore if we tune the parameter $j$ for each vertex operator, there should exist some limit which enforces the LST Green functions on-shell. In that case, the general field theoretic argument suggests that we have LSZ poles:

\begin{equation}
\langle O_1(p_1) \cdots O_n(p_n)\rangle \sim \prod_i \frac{1}{p_i^2+M_i^2} \langle 0|O_1^{(\mathrm{norm})}(p_1) \cdots O_n^{(\mathrm{norm})}(p_n)| 0 \rangle \ , \label{lzhr}
\end{equation} 
where $O_i(p_i)$ is the operator whose corresponding string vertex operator is $\Phi_{j_i,m_i,\bar{m}_i}$ in the $SL(2,\R )_k /U(1)$ part, and $O_i^{(\mathrm{norm})}(p_i)$ is the normalizable (amputated) vertex operator creating a particle from the vacuum. Thus one expects that such poles should appear also from the explicit string calculation. Indeed, it was discussed in \cite{Aharony:2004xn}, for a given $m_i$ and $\bar{m}_i$, which fixes the operator of the LST (from the mass-shell condition of the string theory), that one finds single poles of $j_i$ for particular values. In this context $j_i$ can be seen as a momentum of the dual theory.

In general, it is difficult to separate these LSZ poles of $SL(2,\R )_k /U(1)$ theory from the bulk poles discussed in the last subsections. However, as we will see here, the SRT map provides us with a beautiful framework to do this in quite a manifest way. The first step is to realize that the Liouville part of the SRT map gives the bulk divergence and the group theoretical part encoded in $N^j_{m,\bar{m}}$ gives the LSZ poles. Since $N^j_{m,\bar{m}}$ is attached to every vertex, this is naturally expected, but it is important to see it explicitly and compare with the results of \cite{Aharony:2004xn} as we will do in the following.

{}From its definition \eqref{vertex}\eqref{xbas}, 
\begin{equation}
\Phi_{j,m,\bar{m}} (z) = N^{j}_{m,\bar{m}}\int \frac{d^2\mu}{|\mu|^2} \mu^{-m}\bar{\mu}^{-\bar{m}} \Phi_j(\mu |z) \ ,
\end{equation}
where the normalization factor $N^j_{m,\bar{m}}$ is given by 
\begin{eqnarray}
 \frac{1}{N^j_{m,\bar{m}}} = \frac{1}{\pi}\int d^2s s^{-m+j} \bar{s}^{-\bar{m}+j} e^{s-\bar{s}} \ .
\end{eqnarray}
 The result of the integration is 
\begin{equation}
\int d^2s s^{-m+j} \bar{s}^{-\bar{m}+j} e^{s-\bar{s}} = - \frac{2\pi^2\cos\pi(j-\frac{n+k\omega}{2})}{\sin\pi(2j-k\omega)\Gamma(-j+\frac{-n+k\omega}{2})\Gamma(-j+\frac{n+k\omega}{2})} \ ,
\end{equation}
where 
\begin{equation}
m = \frac{n+k\omega}{2}, \ \ \bar{m} = \frac{-n+k\omega}{2}, \ \ n\in \Z , \ \ \omega \in \Z \ .
\end{equation}
It is easy to see that we can show that 
 the reflection amplitudes \eqref{refwz} are consistent by using this formula and SRT map.

Let us now study the pole structures of $N^{j}_{m,\bar{m}}$. As we mentioned above, this is related to the LSZ pole. To compare with the results in \cite{Aharony:2004xn}, we first note that the index used to label the representations of $SL(2,\R )$ in that reference (let us denote it $j_{AGK}$) is related to ours (let us denote it $j_{RT}$ since our convention agrees with the one used in \cite{rt}) by the Weyl conjugation: $j_{RT} = -1-j_{AGK}$. Thus we have
\begin{equation}
N^{j_{AGK}}_{m,\bar{m}} = \frac{\sin\pi(-2j_{AGK}-2-m-\bar{m})\Gamma(1+j_{AGK}+m)\Gamma(1+j_{AGK}+\bar{m})}{-2\pi\cos\pi(-1-j_{AGK}-m)} \ .
\end{equation}
This amplitude has LSZ pole precisely at 
\begin{equation}
j = M-1,M-2, \cdots >-\frac{1}{2}, \ \ M=\mathrm{min}\{|m|,|\bar{m}|\}, \ \ m,\bar{m} < -\frac{1}{2} \ ,
\end{equation}
which agrees with \cite{Aharony:2004xn} (see eq (2.22) there). On the other hand, in \cite{Aharony:2004xn}, it was also pointed out that there are further LSZ poles in the two-point function at 
\begin{equation}
j = M-1,M-2, \cdots >-\frac{1}{2}, \ \ M=\mathrm{min}\{m,\bar{m}\}, \ \ m,\bar{m} > \frac{1}{2} \ ,
\end{equation}
which do {\it not} appear in $N^{j_{AGK}}_{m,\bar{m}}$. However, as far as $2$-point function concerns, there is no contradiction because the second inserted operator involves the factor $N^{j_{AGK}}_{-m,-\bar{m}}$, which yields the necessary poles because of the momentum conservation.\footnote{In more general amplitudes, these poles arise in the integration over $z_r$ (and the momentum conservation). We will also observe the similar (apparent) asymmetry between $m \to -m$ in the winding violating correlation function in section \ref{sec3-3}.} Therefore we claim that the SRT maps enable us to separate LSZ poles from bulk poles without an ambiguity.

In this way, we have established a holographic meaning of the SRT map: it is a LSZ reduction procedure. We first separate the LSZ poles from the correlation functions of $SL(2,\R )_k/U(1)$ theory and the net interaction is reduced to the Liouville correlation functions. It would be interesting to understand the structure of the remaining Liouville correlation function as a normalized S-matrix (the right hand side of \eqref{lzhr}). This seems feasible because the Liouville correlation function itself gives an S-matrix of a certain noncritical string theory.

\subsection{The violation of winding number conservation}\label{sec3-3}

Our intention here is to extend the correspondence presented in \cite{rt} to the case of WZNW correlation functions that violate the winding number in a generic amount (let us say $\sum _{i=1}^N \omega _i =M$). We will see that it is feasible to do this by using the FZZ prescription.\footnote{G.G. thanks D. Kutasov for a discussion about the details of \cite{FZZ2} at ICTP in April 2001.}  This prescription was reproduced in detail in \cite{mo3}. Then, first, let us discuss how to represent WZNW correlation functions violating the winding number conservation in terms of Liouville correlation functions. In the following, we review the mechanism of violation of the winding number $\omega $ in several related non-rational CFTs.

\subsubsection*{The winding number violation in $AdS_3$ and in the 2D black hole}

The violation of the winding number conservation was first observed by Fateev,
Zamolodchikov and Zamolodchikov (in their quoted unpublished paper \cite{FZZ2}) for the case of the Euclidean version of the Witten's 2D black hole \cite{2DBH}. It was pointed out that the 
conservation of the winding number is not ensured because of the topology of the 
cigar. Furthermore, an interesting prescription was presented to compute 
correlation functions violating the winding number conservation. As the preliminary step of their prescription, the 
concept of ``conjugate representation of the identity operator'' was introduced. This is an operator that basically has the same conformal properties of the identity operator (notice that the conformal dimension of the states in the coset is given by $h_{j,m}^{SL(2)/U(1)} = h_j+m^2/k$); in the case of the 
cigar, the operators $\Phi _{j=-\frac k2,m=\pm \frac k2, \bar{m} =\pm \frac k2}$ are of this sort.

Then, the FZZ prescription continues as follows: First, one takes the $N$-point function one wants to calculate and then one introduces an additional operator (a $N+1^{th}$ operator) with
quantum numbers $j=\pm m=-k/2$ inserted at the point $z_0$;
\begin{equation}
\tilde {\mathcal A}^{WZNW}_{j_1, ... j_N; m_1, ... m_N} = \ \langle \prod _{i=1}^N \Phi _{j_i,m_i,\bar {m}_i} (z_i) \Phi_{j_0=-\frac k2,m_0=\frac k2, \bar{m}_0 =\frac k2}(z_0) \rangle  \ . \label{cardinal}
\end{equation}
In the case one is interested in computing correlation functions that violate the winging number conservation in a generic amount $M$, then $M$ of these conjugate representations of the identity should be included in the correlation functions.

The second step is taking the coincidence limit $z_0 \to z_1$ (let us choose $z_1$ as the position of the vertex operator corresponding to the string state whose winding number will change in the scattering) and then extract {\it ad hoc} the divergence arising in the coincidence limit (this is given by a factor that develops a pole in $m_0 \to \pm k/2$). In order to eliminate the $z_0$-dependence one has to add the overall factor
\[
\prod _{i=1}^N (z _i -z _0)^{m _i}(\bar z _i -\bar z _0)^{\bar m _i}
\]
Besides, when $M$ conjugate representations of the identity are included, this factor would be
\begin{equation}
\prod _{r=1}^M \prod _{i=1}^N (z _i - u_r)^{m _i}(\bar z _i -\bar u _r)^{\bar m _i} \prod _{r<t}^M |u_r -u_t|^{k} \label{gg}
\end{equation}
where the points $u_r$ are those where the additional $M$ operators are inserted (replacing $z_0=u_1$ which corresponds to the case $M=1$).

The third step is to notice that the operator product expansion takes the form
\begin{equation}
\Phi _{j_0,m_0,\bar{m}_0} (z_0) \Phi _{j_1,m_1,\bar{m}_1} (z_1) \sim _{z_0 \to z_1} \sum _{j,m} {\mathcal Q}(j,j_0,j_1;m,m_0,m_1) |z_0-z_1| ^{2(h_j-h_{j_0}-h_{j_1})} \Phi _{j,m,\bar{m}} (z_1) + ... \ ,
\end{equation}
where the Fourier coefficients ${\mathcal Q}(j_1,j_2,j_3;m_1,m_2,m_3)$ are given in terms of the structure constant as follows
\begin{equation}
{\mathcal Q} (j_1,j_2,j_3;m_1,m_2,m_3) = C^H (j_1,j_2,j_3) W^H (j_1,j_2,j_3;m_1,m_2,m_3) {\mathcal R}_k^{-1} (j_1,m_1) \ ,
\end{equation}
while the function $W^H (j_1,j_2,j_3;m_1,m_2,m_3)$ is the group theoretical factor coming from the integration over the $(x_{i},\bar{x}_{i})$-coordinates in changing the base to $(m_i,\bar{m}_i)$-variables ($i \in \{1,2,3\}$) and its explicit form (in terms of hypergeometric functions) is given in \cite{ponjasOpe}. By making use of the functional properties of the structure constant (\ref{quince}) one finds
\begin{eqnarray}
\lim _{j_0 \to -\frac k2} \langle \Phi _{j_0,j_0} (z_0) \Phi _{j_1,m_1} (z_1) ... \Phi _{j_N,m_N}(z_N) \rangle &\sim _{z_0 \to z_1}& \frac {1}{\pi }\frac { B(1-k/2)}{ B(-j_1-k/2)} \tilde {W}_N (j_1,...j_N;m_1,...m_N)  \times \nonumber \\
&& \times \langle \Phi _{-\frac k2-j_1,m_1-\frac k2} (z_1) ... \Phi _{j_N,m_N}(z_N) \rangle   \ ,\label{carita} 
\end{eqnarray}
where, now, the function $\tilde {W}_N (j_1,...,j_N;m_1,...,m_N) $ is given by the integration over the variables $x_a$ when Fourier transforming the expression from the $(x,\bar x)$-basis to the $(m, \bar m)$-basis; including in the integrand the factor required for extracting the $z_0$-dependent divergence mentioned before. For instance, in the case of three-point function ({\it i.e. } $N+1=4$) the integration is 
\begin{eqnarray}
\lim _{x_0 \to \infty} \prod _{a=1}^3 \int &d^2x_a& \prod _{a=1}^3 x_a ^{-1-j_a+m_a} \bar x_a ^{-1-j_a+\bar m _a} |x_{30}|^{4j_3}|x_{21}|^{2j_1+2j_2+2j_3+k}|x_{01}|^{2j_1-2j_2-2j_3-k}  \times \cr 
&&\times |x_{02}|^{-2j_1+2j_2-2j_3-k}|x_0|^{2k}|z-x|^{2j_1+2j_2+2j_3+k}\ , \label{check}
\end{eqnarray}
where $x_{ab}= x_a-x_b$,  $x= \frac {(x_0-x_1)(x_3-x_2)}{(x_2-x_1)(x_3-x_0)}$ and $z = \frac {(z_0-z_1)(z_3-z_2)}{(z_2-z_1)(z_3-z_0)}$, while the factor $|x_0|^{2k}$ (in the limit $x_0 \to \infty$) stands for the regularization required because of the divergence of the Fourier modes $m_0 = \pm k/2$. Besides, the factor $|x-z|^{k+2j_1+2j_2+2j_3}$ comes from solving the Knizhnik-Zamolodchikov equation which has a solution presenting poles at $z=x$. For the details of the explicit form of $\tilde {W}_3 (j_1,j_2,j_3;m_1,m_2,m_3)$ see \cite{mo3}, where the Maldacena and Ooguri (see \cite{FZZ2} for the original computation) find the result being proportional to 
\begin{eqnarray}
\langle  \Phi _{j_1,m_1} (z_1)  \Phi _{j_2,m_2} (z_2) \Phi _{j_3,m_3}(z_3)  \rangle _{violating \ \omega} &\sim &  \frac {\Gamma (1+j_1+j_2+j_3+k/2)}{\Gamma (-j_1-j_2-j_2-k/2)} \prod _{i=1}^3 \frac {\Gamma (-j_i+m_i)}{\Gamma (1+j_i-\bar m _i)} \times \nonumber \\
&& \times (-1)^{m_3-\bar m_3} \pi \ B(j_1) C^H(-\frac k2 -j_1,j_2,j_3) \times \label{wvtpt} \\
&& \times \delta (m_1+m_2+m_3+k/2) \delta (\bar m _1+\bar m _2+\bar m _3+k/2) \ , \nonumber
\end{eqnarray}
Here, the symbol $\sim $ stands because the dependence on $z_i$ was omitted in writing this for simplicity (the details can be found in \cite{mo3}). See also \cite{gn3} for a related calculation.

Then, due to the fact that the starting point of the prescription was an observable with $N+1$ insertions (originally, this is a $N+1$-point function (\ref{cardinal}) that turns out to be the 
$N$-point function (\ref{carita}) after the coincidence limit described above), the conservation law (coming from the integration over the zero-mode of the original vertex-arrange) is $m_1+m_2+m_3+...+m_N+k/2=0$ instead of the usual $m_1+m_2+m_3+...+m_N=0$. Hence, since in the 2D black hole the winding number is given by $\omega _i= m_i+\bar {m} _i$, this procedure enables us to calculate the $N$-point function violating one unit of the winding number in this background. On the other hand, notice that this construction does not imply that the momentum $J^3$ is not conserved. The
conservation law described above comes from the integration over the zero-mode 
of the field $X(z)$ which precisely realizes the $U(1)$ part of the gauged $SL(2,\R )_k/U(1)$ generated by the Cartan current $J^3(z)\sim i\sqrt {k/2} \partial X(z)$. It is just the fact that there are conjugate (additional) representations of the identity operator and the $U(1)$ charge of the accompanying field $X(z)$ is not trivial.

The details of the adaptation of the FZZ prescription described above to the case of string theory in $AdS_3$ can be found in the section 5 of \cite{mo3}. On the other hand, another important remark in \cite{FZZ2} is that (by means of explicit computation) in a given $N$-point function the winding number conservation can be violated in an amount which is bounded by $|\sum_{i=1}^N \omega _i|\leq N-2$. In particular, this means that in a three-point function the winding number can be violated in $-1, \ 0$ or $1$ while in a two-point function the violation can not occur. The reason for the correlation functions with $\sum _{i=1}^N \omega _i>N-2$ to vanish is the regularization factor required to cancel the divergence that is a remnant of the insertion of the conjugate identity operator with $m=\pm k/2$.


In string theory formulated on $AdS_3$ the winding number $\omega $ (as a degree of freedom classifying the states of the spectrum) is understood as follows: Though Anti-de Sitter space is a simply connected manifold, the winding number has to be introduced to describe the Hilbert space. This $\omega $ is certainly not a winding number with a topological origin but it 
is simply due to an energetic reason: In $AdS_3$, the strings are supported by the competition 
between gravity and the action of the $B_{\mu \nu}$ NS-NS field which pushes the strings to the boundary and prevents them from unwrapping in the generic case. However, with enough energy (and under certain circumstances) it is possible for the strings to unwrap and 
change the winding number $\omega $ (by changing $m$ accordingly to ensure that the bulk energy $m + \bar m +k\omega $ is conserved). 
Unlike the 2D black hole, in $AdS_3$ the winding number is not given by 
the sum of the momenta $m$ and $\bar m$ but they represent three independent quantum numbers.

On the other hand, since the eigenvalue of the Cartan generator $J^3_0$ for a generic state in the spectrum of the $SL(2,\R )_k$ WZNW theory is not zero, one could naively argue that in this case, unlike the case of the gauged $SL(2,\R )_k/U(1)$, a state with $m=-k/2$ does not share the conformal properties with the state having $m=0$ and, then, the operator $j=m=-k/2$ 
would not correspond to a conjugate representation of the identity operator $j=m=0$ (because they would have different eigenvalue for $J^3_0$). However, by noticing that in the $SL(2,\R )_k$ WZNW model one has to include the spectral flow symmetry in order to parameterize the whole spectrum (and consequently an infinite new set of states appear in the spectrum), one eventually finds that the state $j=m=-k/2$ with winding number ({\it i.e.} spectral flow parameter) $ \omega =1$ does share the conformal 
properties with the identity $j=m=\omega =0$.
Hence, in the case of $AdS_3$ string theory the state with winding number $\omega =1$ is the one that corresponds to the conjugate representation of the identity (this was called {\it spectral flow operator} in \cite{mo3}). Hence, the prescription for computing violating winding amplitudes in $AdS_3$ turns out to be similar to the one for the 2D black hole and one finds analogous results and the same bound for the violation of $\omega $ (see appendix D of \cite{mo3}).

\subsubsection*{The SRT dictionary and the violation of winding number}

At this point, the question arises as to whether there is a description of the WZNW correlation functions that violate the winding number conservation in terms of Liouville correlation functions ({\it i.e.} in an analogous way as (\ref{rt}) describes such a map for the winding number conserving correlation functions). In order to undertake the task of including the winding number violating correlation functions in the framework, we have to start with the following simple observation: From the relation between the quantum numbers $j_i$ and $\alpha _i$ in (\ref{rt}), one notices that a correlation function of the form (\ref{cardinal}) which includes a $N+1^{th}$ operator with momentum $j=-\frac k2$, corresponds to the insertion of the identity operator $V_{\alpha_0 =0}\sim 1$ in the Liouville theory side. Then, correlation functions of the form (\ref{cardinal}) are actually proportional to Liouville correlation functions as
\begin{equation}
\lim _{\alpha _0 \to 0} \prod _{l=1}^{N-1} \int d^2v_l \langle V_{\alpha _0} (z_0) \prod _{i=1}^N V_{\alpha _i}(z_i) \prod _{l=1}^{N-1} V_{\alpha _{1,2}}(v_l)\rangle \ . \label{c43}
\end{equation}
Then, since the Liouville structure constants $C^{L}(\alpha _1, \alpha _2, \alpha _3)$, like their analogues  $C^{H}(j _1,j _2,j _3)$, satisfy
\begin{equation}
\lim _{\alpha_0 \to 0} C^L (\alpha_0,\alpha_1,\alpha) = R_b(\alpha ) \delta (\alpha -\alpha_1) \ ,
\end{equation}
one notices that the OPE (\ref{carita}) in the WZNW theory side leads to a result proportional to
\begin{equation}
\langle  \prod _{i=1}^N V _{\alpha _i} (z_i) \prod _{l=N+1}^{2N-1} V_{-\frac {1}{2b}} (z_l)   \rangle   \label{carita2}
\end{equation}
in the Liouville theory side. Notice that in (\ref{carita2}), there are $2N-1$ Liouville vertex operators $V_{\alpha}$ instead of $2N-2$. Then, the correspondence between (\ref{carita}) and (\ref{carita2}) suggests a way to extend the Stoyanovsky-Ribault-Teschner dictionary in such a way that the winding violating processes can be represented also in terms of Liouville correlation functions.

Following the analysis presented above and extending to the $M$ winding violating case, one concludes that the $N$-point WZNW correlation functions violating the total winding number conservation in an amount $\sum _{i=1}^N \omega _i=M$ can be written as $2N+M-2$-point functions in Liouville field theory (in the case $M=0$ one recovers the particular case (\ref{rt}) valid for the winding preserving process). This is due to the fact that each spectral flow operator $\Phi _{j=-\frac k2}$ that needs to be inserted in the WZNW side in order to change (in one unit) the winding number of a given $SL(2,\R )$-state corresponds to an identity operator $V_{\alpha =0} \sim 1 $ in the Liouville theory side and then one observes that
\begin{equation}
\langle \prod _{r=1}^{M} \Phi _{-\frac k2} (u_r) \prod _{i=1}^N \Phi_{j_i} (z_i) \rangle \propto
\prod _{l=1}^{N+M-2}\int d^2v_l \langle \prod _{l=1}^{N+M-2} V_{\alpha _{1,2}} (v_l)  \prod _{i=1}^{N} V_{\alpha _i} (z_i)   \rangle \ .
\end{equation}
For instance, the three-point scattering amplitudes (violating winding number with $M=1$) in $AdS_3$ would be given in terms of five-point Liouville correlation functions (extracting the appropriate divergent factor coming from the coincidence limit of spectral flow operator and the evaluation at $m = \bar m = \pm k/2$; see below). Thus, according to this digression, only those Liouville correlation functions involving $N+n$ vertex operators with $n$ of them being degenerate states with momenta $\alpha _{1,2}$ and $N-2\leq n \leq 2N-2$ can be thought as WZNW correlation functions (violating the winding number conservation in an amount $\sum _{i=1}^N \omega _i= n-(N-2)$.)

Hence, the closed formula realizing the map would be
\begin{eqnarray}
\langle\prod _{i=1}^N \Phi_{j_i,m_i,\bar m_i}(z_i)\rangle_{violating \ \omega} \ = \frac {\pi^3b}{2^{2N+2M-1}} \lim _{p_1 \to -k/2} \ ... \lim _{p_M \to -k/2}  \prod _{r=1}^{M} (p_r+\frac k2) \prod _{r<t}^{M} |u_r-u_t|^{2(p_r +p_t +k)} \nonumber \times
\end{eqnarray}
\begin{eqnarray}
\times \prod _{i=1}^{N} N^{j_i}_{m_i,\bar m_i} \prod _{r=1}^{M} N^{-k/2}_{p_r,p_r} \ \prod _{r=1}^{M} \prod _{i=1}^{N} (z_i-u_r)^{2m_i p_r/k+m_i+p_r+k/2} (\bar z _i-\bar u _r)^{2\bar m _i p_r/k+\bar m _i+p_r+k/2} \times \nonumber \\
\times \prod _{i<j}^{N}(z_i-z_j)^{2m_im_j/k+m_i+m_j+k/2} (\bar z _i-\bar z _j)^{2\bar m _i \bar m _j/k+\bar m _i +\bar m _j + k/2} \prod _{l=1}^{N+M-2} \int d^2v_l \prod _{l<t}^{M+N-2}|v_l -v_t|^{k} \times \nonumber \\
\times  \prod _{i=1}^{N} \prod _{l=1}^{N+M-2} (z _i- v _l)^{-m _i-k/2} (\bar z _i- \bar v _l)^{-\bar m _i-k/2} \prod _{r=1}^{M} \prod _{l=1}^{N+M-2} |u_r-v_l|^{-2p_l-k} \times \nonumber \\
\times  \ \langle\prod _{i=1}^N V_{\alpha _i} (z_i) \prod _{l=1}^{N+M-2}V_{-\frac {1}{2b}}(v_l)\rangle \ \delta (\sum _{i=1}^N m_i -Mk/2) \delta (\sum _{i=1}^N \bar m _i -Mk/2) \ ,\label{postaa}
\end{eqnarray}
where we have denoted $v_l = z_{N+M+1+l}$ for $0<l\leq N+M-2$ while the $u_r$ correspond to the points where the $M$ spectral flow operators are inserted (analogues to the $z_0$ in (\ref{c43})). The factors of the form $(z_i -z_j)^{2m_i m_j /k}$ come from the fact that we are computing correlation functions in the gauged $SL(2,\R )/U(1)$, and these correspond to the contraction of two operators $e^{i\sqrt {\frac 2k}m_i X(z_i)}e^{i\sqrt {\frac 2k}m_j X(z_j)}$.

The factor $\prod _{r=1}^M (p_r+k/2)$ in the expression above stands because of the regularization required in order to cancel the pole arising in the limit $p_r\to -k/2$. This {\it regularization} is explained in \cite{mo3} by introducing the infinite volume factor $1/V_{conf}$ which is attached to each spectral flow operator. In terms of the ($m,\bar m$)-basis, such a divergence that arises in the limit $p_r \to \pm k/2$ is due to the factor $N^{-k/2}_{p_r,p_r}$. Thus, the regularization factors $(p_r+k/2)$ precisely cancel such a divergence in the right hand side of the last equation; namely
\[
\lim _{p_r\to -k/2}(p_r+k/2)N^{-k/2}_{p_r,p_r} = \lim _{p_r \to -k/2}-\frac {(p_r+k/2)\sin(\pi(k+2p_r))\Gamma^2 (k/2+p_r)}{2\pi \cos(\pi (k/2+p_r))} = -1 \ .
\]
Besides, notice that this also verifies that the factor (\ref{gg}) is the appropriate one in order to cancel the $u_r$-dependence coming from the insertion of the $M$ additional spectral flow operators $\Phi _{-\frac k2,- \frac k2,- \frac k2}$. This is observed from the fact that in the limit $p_r\to -k/2$ the $u_r$-dependent factors cancel (see equation above). Then, one finally obtains
\begin{eqnarray}
\langle \prod _{i=1}^N \Phi_{j_i,m_i,\bar m_i}(z_i)\rangle_{violating \ \omega}  \ = \ \cr
= \frac {(-1)^M\pi^3b}{2^{2(N+M)-1}} \prod _{i=1}^{N} N^{j_i}_{m_i,\bar m_i}\prod _{i<j}^{N}(z_i-z_j)^{2m_im_j/k+m_i+m_j+k/2} (\bar z _i-\bar z _j)^{2\bar m _i \bar m _j/k+\bar m _i +\bar m _j + k/2} \times \cr
\times  \prod _{l=1}^{N+M-2} \int d^2v_l \prod _{l<t}^{M+N-2}|v_l -v_t|^{k} \prod _{i=1}^{N} \prod _{l=1}^{N+M-2} (z _i- v _l)^{-m _i-k/2} (\bar z _i- \bar v _l)^{-\bar m _i-k/2} \times \nonumber \\
\times \prod _{r=1}^{M} \prod _{l=1}^{N+M-2} |u_r-v_l|^{-2p_l-k} \times \nonumber \\
\times  \ \langle\prod _{i=1}^N V_{\alpha _i} (z_i) \prod _{l=1}^{N+M-2}V_{-\frac {1}{2b}}(v_l)\rangle \ \delta (\sum _{i=1}^N m_i -Mk/2) \delta (\sum _{i=1}^N \bar m _i -Mk/2) \ . \label{postaaaa} 
\end{eqnarray}
Hence, the formula (\ref{postaa}) generalizes the SRT correspondence (\ref{rt}) since it enables us to describe winding violating processes in WZNW model in terms of Liouville correlation functions\footnote{In \cite{rt} it was noted that V. Fateev derived a formula relating winding violating correlation functions in WZNW model with Liouville correlation functions in his unpublished paper.}.

Let us focus on the winding number violating $(M=\pm 1)$ three-point function as the first nontrivial example. We first note that the nontrivial group theoretical factor ($m,\bar{m}$ dependence) is given by (see \eqref{wvtpt})
\begin{equation}
\tilde{W}_N(j_1,j_2,j_3,m_1,m_2,m_3) \propto \prod_{i=1}^{3} \frac{\Gamma(-j_i+m_i)}{\Gamma(1+j_i-\bar{m}_i)} \ .
\end{equation}
On the other hand, this factor is nothing but the product of the normalization factor $N^j_{m,\bar{m}}$ appearing in \eqref{postaaaa}
\begin{equation}
\prod _{i=3}^{N} N^{j_i}_{m_i,\bar m_i} = \prod_{i=1}^{3} \frac{\Gamma(-j_i+m_i)}{\Gamma(1+j_i-\bar{m}_i)} \ ,
\end{equation}
so we have successfully separated the group theoretical factor again. We also notice that the leading singularity of the Liouville five-point function in the $v_1 \to z_1 \sim z_0$, $v_2 \to z_1\sim z_0$ limit\footnote{In the case of $m=\frac{k}{2}$, the divergent contribution comes from this part because $\lim_{p_r \to +k/2} N^{-k/2}_{p_r,p_r}$ is now finite. Instead the integration over $v_i$ develops a diverging factor when approaching $z_0$.} gives a factor proportional to $C^H(-\frac{k}{2}-j_1,j_2,j_3)$ by a direct calculation, agreeing with the WZNW model computation \eqref{wvtpt}.

For completeness, we would like to review the mechanism leading to the violation of the winding number conservation in other (though closely related) non-rational CFTs. In the sine-Liouville field theory, which was conjectured to be dual to $SL(2,\R)_k /U(1)$ coset model, the mathematical reason for the violation of winding conservation is more evident than in the case of the $SL(2,\R)_k /U(1)$ coset model (even though the geometrical picture for such a mechanism is quite clear in the spacetime interpretation). This is due to the fact that the winding 
number in sine-Liouville theory is explicitly broken by the interaction term (which, unlike the WZNW theory, explicitly depends on the field $X(z)$). The sine-Liouville potential is chiral 
in the sense that a distinction between right-moving-modes and left-moving-modes of the field $X(z)$ manifestly appears in the non-trivial part of the action. Fukuda and Hosomichi showed in \cite{fh} that one can rewrite the interaction term (see (\ref{sfzz}) below) as 
two exponential terms and these terms turn out to represent two different screening 
charges of the model. Then, the correlation functions can be computed by using 
an amount $n$ and $m$ of both screenings respectively. The computation that uses the same amount of both ($m=n$ in the notation of \cite{fh}) conserves the winding number while the one employing a different amount violates the winding number in $\sum_{i=1}^N \omega _i =n-m$ units.

At this point one could feel tempted to reproduce this procedure in the case of the $SL(2,\R )_k$ WZNW model where, as well as in the sine-Liouville model, one finds two different screening charges. However, in the WZNW model both 
screening charges are similar with respect to the winding charge; thus (as 
proven in \cite{gn3}) all the computations using a generic amount of screening 
of any kind in the case of the WZNW theory agree and the violation of winding comes 
from a rather different aspect.

Fukuda and Hosomichi also showed that the violation (at least up to the 
level of three-point function) is bounded by $\sum _{i=1}^{N=3} \omega \leq N-2$ (this seems consistent with the upper bound $N-2$ in the WZNW model). This upper bound also appears in the context of the $\mathcal{N}=2$ superconformal $SL(2,\R )_{k+2}/U(1)$ model. In the case one is dealing with a supersymmetric model ({\it i.e.} the model with $\mathcal{N}=2$ SCA superconformal symmetry) 
one finds two basic aspects which can lead to a simpler understanding of the winding number violation. First of all, the SCA algebra also presents the spectral
flow symmetry and under this transformation the local supercurrents $G^+(z)$ and $G^-(z)$ change in such a way that their modes are shifted. On the other hand, in the supersymmetric case the 
calculation of $N$-point functions involves $N-2$ states written in the picture $0$ and two states in the picture $-1$. In order to write the operators in the picture $0$ one 
should insert $N-2$ supercurrents in the correlation functions and then, under the 
application of the spectral flow symmetry, some of these $N-2$ currents can 
change (one of the contribution of the complete expression containing $N-2$ supercurrents insertions) and the violation of the winding 
number can be understood from this fact (for the details, see \cite{little}). This mechanism 
led Giveon and Kutasov to remark the parallelism between the supersymmetric case and what 
happens in the bosonic case ({\it i.e.} the fact that the winding number can be 
violated up to $N-2$ units).\footnote{The supersymmetry version of the FZZ duality was proven in \cite{Hori:2001ax}. From this gauged linear sigma model viewpoint, the violation of winding number conservation is due to instanton effects.}

\subsection{The $k\to 0$ limit: The homogeneity limit of the tachyon field}\label{sec3-4}

As we commented in the introduction, the WZNW model formulated on $SL(2,\R )_k$, in addition to its natural application to string theory on $AdS_3$ and on the 2D black hole, is also connected to the tachyon physics in string theory through the FZZ conjecture. These aspects are reviewed in the following paragraph as an introduction to this subsection.

\subsubsection*{Preliminary: On the FZZ conjecture} 

Let us begin with the Fateev-Zamolodchikov-Zamolodchikov conjecture \cite{FZZ2}, which will be of primary importance in this section. This conjecture states the equivalence between $N$-point correlation functions in the two-dimensional black hole or $SL(2,\R )_k/U(1) $ coset model, and $N$-point functions in the sine-Liouville conformal model. The latter model is described by the following action
\begin{equation}
S_{FZZ} =\frac {1}{2\pi} \int d^{2}z\left( \partial \varphi\bar{\partial}\varphi   +   \partial X\bar{\partial}X       - \frac {1}{2\sqrt {k-2}} R \varphi \right) + \lambda \int d^2z \ T  \label{sfzz}
\end{equation}
with the interaction
\[
T(z,\bar z) =  e^{-\sqrt {\frac {k-2}{2}}\varphi } \ \cos (\sqrt {\frac k2} (X_L(z)-X_R(\bar z)))
\]
defining an exact conformal background that can be thought as a deformation of the $c=1$ matter model \cite{kkk}. The central charge is given by $c= 2+\frac{6}{k-2}$ and consistent with the FZZ conjecture. The primary fields we are going to deal with are represented by operators of the form
\begin{equation}
V_{\alpha ,p_L,p_R}^{FZZ} (z,\bar z)= V_{\alpha }(z,\bar z) \ e^{i\sqrt{2} (p_L X_L (z)+p_R X_R (\bar z))}   \ .\label{vvv}
\end{equation} 
The three-point correlation functions involving these fields were computed in \cite{fh} and were shown to be consistent with the FZZ conjecture (see also \cite{daniel}).

According to the FZZ conjecture, the sine-Liouville model (\ref{sfzz}) is dual to the string theory formulated on the Euclidean black hole, {\it i.e.} the $SL(2,\R )_k/U(1)$ coset model. The spectrum of this model, following the notation introduced before, is given by operators having conformal dimension $\tilde {\Delta} _{\alpha , p_L}$, given by
\begin{equation}
\tilde {\Delta } _{\alpha ,p_L} = - \alpha (\alpha +b) +p_L^2  \label{peso}
\end{equation}  
and the analogous expression for $p_L \to p_R$. It is convenient to introduce the following notation
\begin{equation}
m = \sqrt {k} \ p_L , \ \ \bar m = \sqrt {k} \ p_R , \ \ \alpha =bj, \ \ b^{-2}=k-2  \ ,\label{vvvv}
\end{equation}
where the range of quantum numbers is defined by the grid
\begin{equation}
m+\bar m = \omega  k \ \ \ \ , \ \ m-\bar m = n 
\end{equation}
with $(\omega , n)$ being a pair of integers. The notation in (\ref{vvvv}) is convenient for making the map between states of both sine-Liouville and the WZNW theory clear enough. Notice that, with this nomenclature, the map between states is realized simply by $\tilde{\Delta } _{\alpha,p_L} = h_{j,m}^{SL(2)/U(1)}$.

\subsubsection*{The $k\to 0$ limit of sine-Liouville field theory}

A preliminary remark that turns out to be important for our discussion here is that the action $S_{FZZ}$ actually coincides with the Liouville action $S_{L}$ at the point $k=0$ of the space of parameters once we make the identification $b^{-2}=k-2$. This is simply verified by noticing that, at $k=0$, the background charges of both actions agree and the sine-Liouville potential ({\it i.e.} the tachyon field) becomes a single exponential term $T(z,\bar z)\sim e^{-i\phi (z,\bar z)}$. Let us also notice that the second screening operator $S_-$ of Liouville theory decouples because $\mu _- $ actually vanishes for $b^2=-1/2$. Finally, the central charge of the sine-Liouville theory is given by $c = 2+ \frac{6}{0-2} = -1$, which agrees with the Liouville theory at $b^2=-1/2$ $+$ a free boson: $c = (1-6\cdot\frac{1}{2}) + 1 = -1$.

The equivalence between actions is a suggestive fact; however, as mentioned, since the theories are strongly coupled for $k<2$, inferring properties from the coincidence of both actions is not sufficiently evident. The equivalence between Liouville theory and sine-Liouville theory at $k=0$, if any, should be tested at the level of correlation functions. Here we will argue that, by assuming the FZZ conjecture, the sine-Liouville model actually coincides with the Liouville theory $\times$ a free $U(1)$ boson $\partial X(z)\bar{\partial}X(z)$ in the limit $k\to 0$ even at the quantum level.
This limit proves that the ``homogeneity limit'' analyzed in \cite{takayanagi} is perfectly consistent.

In order to prove this, it is convenient to employ the dictionary (\ref{rt}). It is feasible to show that the Stoyanovsky-Ribault-Teschner dictionary, in the limit $b^2 \to -1/2$, turns out to give a direct equivalence between correlation functions in both Liouville and sine-Liouville theories. Once FZZ conjecture is assumed to hold for any value of $k$, proving the coincidence of the correlation functions both in the Liouville theory and the sine-Liouville theory in the limit $b^{-2} \to -2$ is simply equivalent to observing a series of remarkable facts in the SRT map in this limit:

\begin{description}
	\item[a)] The exponent of the function $\Theta _N(\mu | z) $ connecting Liouville correlation functions and WZNW correlation functions vanishes when it is evaluated at $k=0$ (recall (\ref{caballo})). In other words, nontrivial dependence (other than the Liouville correlation function) on $z_{N+1} \dots z_{2N-2}$ in \eqref{rt} vanishes if we set $m_i = \bar{m}_i = 0$.
	\item[b)] Since we are taking a limit such that $R = \sqrt k $ goes to zero ({\it i.e.} the asymptotic radius of the cigar), it is enough to observe what  happens with the modes $m=\bar m = 0$ on the cigar (for instance, in order to make the equation (\ref {peso}) to make sense). From the point of view of sine-Liouville model, the dual radius $\tilde {R} \sim 1/\sqrt{k}$ of the cylinder goes to infinity and the states with finite momentum $p=\frac {m}{\sqrt k}$ (keeping $p$ fixed) decouple generating a $U(1)$ factor $\sim e^{i\sqrt {2} p X}$ in the correlation functions.
	\item[c)] Then, by using the formulae
\[
\Gamma (2x) = 2^{2x-1} \pi ^{-1/2} \Gamma (x)  \Gamma (x+\frac 12) \ \ , \ \ \ \sin (\pi x) = \pi \Gamma ^{-1} (x) \Gamma ^{-1} (1-x)
\]
we eventually obtain 
\begin{eqnarray}
\langle \prod _{i=1}^N \Phi _{j_i,m _i,\bar {m}_i} (z_i)  \rangle &=&  (\pi ^2 \ \mu _+/M) ^{\sum_{i=1}^N j_i +1} \prod _{i=1}^N {\mathcal R}_0 (j_i, 0) \prod _{t=1}^{N-2} \int d^2 v_t \ \langle \prod _{i=1}^N V _{bj_i} (z_i) \prod _{t=1}^{N-2} V_{b} (v_t)  \rangle \times \nonumber \\
&& \times \delta (\sum_{i=1}^N p_{R_i}) \delta (\sum_{i=1}^N p_{L_i}) \ ,\label{pros}
\end{eqnarray}
where we have absorbed a $j$-independent factor in the normalization of vertex $\Phi _j$ and a global $N$-independent factor in the definition of the inner product.
\end{description}
Notice that the exponent of $\mu _+$ in the last formula precisely coincides with the one expected from KPZ scaling in sine-Liouville theory and, consequently, enables one to identify the couplings of both CFT's as being $\lambda \sim_{k=0} \pi^2 \mu _+ /M$. Finally, we have also used 
\[
{\mathcal R}_{k=0}(j , m=0) = 2^{2j+1} \Gamma (-j)/\Gamma(1+j) \ ,
\]
which is the reflection coefficient of WZNW model at $k=0$ (and of course, the reflection coefficient of the Liouville theory at $b^2=-1/2$), whose presence is perfectly in agreement with the representation used in (\ref{vertex}). The existence of the reflection amplitudes for each vertex operator is ultimately attributed to the fact that the sign of the background charge is different between sine-Liouville theory and Liouville theory in this convention.

We also observe in (\ref{pros}) that, besides the $n_+$ integrals involved in the Liouville correlation functions, we get $N-2$ additional integrals over the variables $v_t$. This is consistent with what one would expect and it is explained by the following two observations: First, notice that in the limit $k \to 0$ the Liouville degenerate operator $\int d^2v V_{\alpha _{1,2}}(v)$ turns out to coincide with the screening charge $\int d^2v V_b(v)\sim S_+$; on the other hand, when comparing (\ref{pocho}) with (\ref{clot}), we find $n_+ - s= (N-2)$. Accordingly, the additional $N-2$ integrated operators $\int d^2v_t V_b(v_t)$ do have to be present in order to compensate the amounts of screening charges of both theories, which guarantees that the correlation functions do not vanish. This is particularly true in the Euclidean signature of $k<2$ models because the zero mode integration gives not a pole of $\Gamma$-function but a $\delta$-function enforcing the strict (Liouville) momentum conservation. The necessity of ``extra screening operators" becomes subtler when one considers Lorentzian signature after the Wick rotation. One should note, however, the screening integral is just a cosmological constant operator and seems harmless in the general structure of the correlation functions.

 Related to this point, it was recently pointed out in \cite{Zamolodchikov:2005fy,Kostov:2005kk} (see also \cite{schomerus,Balasubramanian:2004fz} in the context of the time-like Liouville theory), that the higher point functions than two-point functions in the imaginary $b$ theory (in the Euclidean theory such as the minimal model) are different from the analytic continuation from the real $b$ theory. Although this might potentially seem to invalidate our formula \eqref{pros}, it is actually not. The reason is that the SRT map is derived {\it both} when $b$ is real and imaginary.\footnote{In fact, Stoyanovsky proposed his map in the $SU(2)_k$ WZNW model corresponding to an imaginary $b$.} Therefore, as long as we interpret the both sides of \eqref{pros} in the same manner ({\it i.e.} if we interpret the correlation functions of the sine-Liouville theory in the Euclidean (Lorentzian) manner, we should do so in the Liouville theory), we obtain consistent correlation functions.

 Then, the integrals over the $v_t$ variables are nothing more than screening insertions in Liouville correlation functions, which is required to satisfy the conservation laws from the integration over the zero-mode of the field $\varphi $. Hence, for $k=0$ the map (\ref{rt}) turns out to be an identity between $N$-point functions in Liouville theory and $N$-point functions in the 2D black hole. 
This completes the proof of the equivalence presented in the introduction.

\[
\]

\section{Discussion and Conclusion}\label{sec4}
In this note, we have discussed the physical applications of the SRT map to string theory in the noncompact curved background. In particular, we have shown that the SRT dictionary encodes several important physical aspects of the pole structure of the string scattering amplitudes in $AdS_3$ space and two dimensional black holes. The structure is perfectly consistent with the $AdS_3/CFT_2$ correspondence and holographic dual description of the LST. Furthermore, we have proposed a formula describing the winding violating correlation functions in $SL(2,\C)_k/SU(2)$ WZNW model in terms of Liouville correlation functions, extending the SRT map.  Finally, we have studied the WZNW correlation functions in the limit $k \to 0$ and shown that they agree with those of the Liouville field theory. This result makes contact with recent studies on the dynamical tachyon condensation in closed string theory. Our results are consistent with the FZZ conjecture and proved its validity at $k=0$.

There would be still other interesting applications of the SRT map. For example, in his recent paper \cite{Zamolodchikov:2003yb}, Zamolodchikov introduced the higher equations of motion (one-to-one corresponding to the BPZ equations) in the Liouville theory, and pointed out that these equations might be important in the minimal string theory (or minimal gravity). As we pointed out in \cite{Bertoldi:2004yk}, the derivation of Zamolodchikov's higher equation of motion possesses  some model independent feature, and can be generalized to other noncompact CFTs. Indeed, \cite{Bertoldi:2004ye} derived the analogous equations in $SL(2,\C)/SU(2)$ WZNW model. 

The geometrical relation between the Zamolodchikov equations in the Liouville theory and their counterpart in WZNW model was studied in \cite{Bertoldi:2004yk} in the classical branch, where we have observed that the hidden Liouville geometry of the WZNW model plays a crucial role. The central question is whether the SRT map yields yet another correspondence between these two equations in these two different theories. Preliminary observations state
\begin{itemize}
	\item The map of the parameter \eqref{mapl} suggests that the logarithmic operator of the Liouville theory be mapped to logarithmic operator of the WZNW model in the $\tilde{j}^-_{m,n} = \frac{m-1}{2}-\frac{n}{2}(k-2)$ branch:
\begin{eqnarray}
\tilde{\alpha}_{m,n} = \frac{m+1}{2}b - \frac{(n-1)}{2b} = b\tilde{j}^{-}_{m,n} + b + \frac{b^{-1}}{2} \ .
\end{eqnarray}
	\item The Zamolodchikov coefficients in both theory take the similar functional dependence on the parameter after the map \eqref{mapl}.
\end{itemize}
It would be very interesting if one can derive the other Zamolodchikov operator valued equation from the other one by using the SRT map.

Regarding the applications of the SRT map as realizing string scattering amplitudes in $AdS_3$, there are still some open questions. One of these is the question about the constraints that have to be imposed on the external momenta of a given scattering process for the amplitude to be well defined. From the viewpoint of the $AdS_3 /CFT_2$ correspondence, the $N$-point string amplitudes in $AdS_3$ have a natural physical interpretation only if the momenta of the incoming (and outgoing) strings satisfy the relation
\begin{equation}
\sum_{i=1}^N j_i > 3-N-k  \label{xxxx}
\end{equation}
which, in terms of the Liouville theory, means
\begin{equation}
\sum_{i=1}^{2N-2} \alpha _i > b
\end{equation}
with $\alpha _i =-1/2b$ for $N<i<2N-2$. However, it does not seem to be simple to explain such a bound within the context of Liouville theory. In the particular case of the two-point function ($N=2$) the above condition turns out to be
\begin{equation}
\frac {1-k}{2}<j \ \ , \ \ \ 2<k \label{x}  \ ,
\end{equation}
which corresponds to the (improved) unitarity bound \cite{mo1}. Moreover, the unitarity of the free string spectrum in $AdS_3$ requires
\begin{equation}
\frac {1-k}{2}<  j < -\frac 12 \label{xx}  
\end{equation}
and the upper bound of (\ref{xx}) precisely corresponds to the Seiberg bound in the Liouville theory side:
\begin{equation}
\alpha  < \frac Q2 \ .  \label{xxx}  
\end{equation}
However, no such a clear identification holds for higher point functions ({\it i.e.} for $N>2$ in (\ref{xxxx})).

Another interesting puzzle is the question about the meaning of the constraints required for the OPE of the dual CFT to factorize. In \cite{mo3}, it was proven that, after integrating over $z_i$, only some of the whole set of integrated four-point functions in $AdS_3$ can be factorized with a clear interpretation of the intermediate states. These {\it special} four-string amplitudes are those that satisfy certain constraints, which are even more restrictive than (\ref{xxxx}). Then, the second open question would be the interpretation of such constraints from the Liouville viewpoint. In the framework of the $AdS_3/CFT_2$ correspondence, those constraints have a clear meaning as the ones required to avoid non-local effects in the boundary of $AdS$. Their interpretation from the point of view of the integrated $6$-point function in Liouville field theory could be useful as well. Furthermore, it should be also interesting to understand what are the interpretation of those as necessary conditions for the corresponding Liouville-string amplitudes to be well defined.

\section*{Acknowledgements}
Gaston Giribet thanks Simeon Hellerman, Sameer Murthy and Ari Pakman for useful conversations on sine-Liouville field theory and related topics. He also thanks the Abdus Salam International Center for Theoretical Physics for the hospitality. Yu Nakayama would like to express his gratitude to Marco Matone and Padova university for their hospitality during his stay, where the early stage of this work was performed. The authors thank Sylvain Ribault for interesting comments. They are also very grateful to Tadashi Takayanagi for discussions on this subject.

Gaston Giribet is supported by CONICET and UBA, Argentina.
Yu Nakayama is supported in part by JSPS
Research Fellowships for Young Scientists.


\end{document}